\documentstyle[aps,prl,twocolumn,floats,epsf]{revtex}
\begin{document}
\wideabs{

\title{Evidence for a Goldstone Mode in a Double Layer Quantum Hall System}
\author{I.B.~Spielman$^1$, J.P.~Eisenstein$^1$, L.N.~Pfeiffer$^2$ and K.W.~West$^2$}
\address{$^1$California Institute of Technology, Pasadena, CA 91125}
\address{$^2$Bell Laboratories, Lucent Technologies, Murray Hill, NJ 07974}
 
\date{\today}

\maketitle \begin{abstract} The tunneling conductance between two parallel 2D electron
systems has been  measured in a regime of strong interlayer Coulomb correlations.  At
total Landau  level filling $\nu_T=1$ the tunnel spectrum changes qualitatively when the
boundary  separating the compressible phase from the ferromagnetic quantized Hall state
is  crossed.  A huge resonant enhancement replaces the strongly suppressed  equilibrium
tunneling characteristic of weakly coupled layers. The possible  relationship of this
enhancement to the Goldstone mode of the broken symmetry  ground state is discussed.
\end{abstract}

}

When two parallel two-dimensional electron systems (2DES) are sufficiently close 
together, interlayer Coulomb interactions can produce collective states which have no 
counterpart in the individual 2D systems \cite {book,suen,jpe1}.  One of the simplest,
yet most interesting,  examples occurs when the total electron density, $N_T$, equals
the degeneracy $eB/h$ of a  single spin-resolved Landau level produced by a magnetic
field $B$. In the balanced case  (i.e. with layer densities $N_1$=$N_2$=$N_T/2$), the
Landau level filling factor of each layer  viewed separately is $\nu =hN_T/2eB$=$1/2$. 
If the separation $d$ between the layers is large,  they behave independently and are
well described as gapless composite fermion liquids.   No  quantized Hall effect (QHE)
is seen.  On the other hand, as $d$ is reduced, the system  undergoes a quantum phase
transition\cite{boebinger,macdonald,santos} to an incompressible state  best described
by the total filling factor $\nu_T$=$1/2+1/2$=$1$.  A quantized Hall plateau now 
appears at $\rho_{xy}=h/e^2$.  Both Coulomb interactions and interlayer tunneling
contribute to the strength of this QHE  but there is strong evidence from experiment
\cite {jpe1,murphy} and theory \cite{book,chak}  that the incompressibility  survives
in the limit of zero tunneling.  This remarkable collective state exhibits a broken 
symmetry\cite{fertig,wen,yang}, spontanteous interlayer phase coherence,
and may be viewed as a new kind of easy-plane ferromagnet. 
The  magnetization of this ferromagnet exists in a pseudospin space; electrons in one
layer are  pseudospin up, while those in the other layer are pseudospin down.  Numerous
interesting  properties are anticipated, including linearly dispersing Goldstone
collective modes  (i.e. pseudospin waves), a finite temperature Kosterlitz-Thouless
(K-T) transition,  dissipationless transport for currents directed oppositely in the
two layers, and bizarre  topological defects in the pseudospin
field\cite{fertig,wen,yang,stern}.  To date, most experimental  results on this system have
derived from electrical transport  measurements\cite{jpe1,boebinger,murphy,lay,sawada}
although recently an optical  study has been reported\cite{manfra}.

In this paper we report a new study of the double layer $\nu_T$=1 ferromagnetic quantum
Hall  state, and its transition at large layer separation to a compressible phase,
using the method  of tunneling spectroscopy.  Earlier experiments have shown that there
is a very strong  suppression of the equilibrium tunneling between two widely separated
parallel 2DESs at  high magnetic field\cite{jpe2,brown}.  This suppression is a result
of the energetic penalty  accompanying the rapid injection (or extraction) of an
electron into the strongly  correlated electron system produced by Landau
quantization.  Other than a small  downward shift in energy produced by the excitonic
attraction of a tunneled electron and  the hole it leaves behind in the source
layer\cite{jpe3}, the measured tunneling spectrum is  simply a convolution of the
spectral functions in the individual layers.  Here we show that  when the layers are
close enough together to support the bilayer $\nu_T$=1 QHE state, this  is no
longer the case.  The strong suppression is replaced by a huge resonant  enhancement of
the tunneling.  The appearance of this resonance suggests the existence  of a soft
collective mode of the double layer system which enhances the ability of electrons to
tunnel. This may well be the predicted\cite{fertig,wen} Goldstone mode of the broken  symmetry
ferromagnetic ground state at $\nu_T$=1.

The samples used in this experiment are $\rm{GaAs/Al_xGa_{1-x}As}$ double quantum well
(DQW)  heterostructures grown by molecular beam epitaxy (MBE).  Two 180{\AA} GaAs
quantum  wells are separated by a 99{\AA} $\rm{Al_{0.9}Ga_{0.1}As}$ barrier layer.  The
DQW is embedded  in thick  $\rm{Al_{0.3}Ga_{0.7}As}$ layers which contain Si doping
sheets set back from the GaAs wells  sufficiently far to produce 2D electron gases in
each well with nearly equal electron  densities of $5.4\times 10^{10}cm^{-2}$.  A
square mesa, $250\mu m$ on a side, is  patterned using standard  photolithography. 
Gate electrodes deposited above and below this mesa allow control  over the densities
$N_{1,2}$ of the two 2D layers.  The low temperature mobility of the as-grown sample
is $7.5 \times 10^5cm^2/Vs$ but this drops to $2.5 \times 10^5cm^2/Vs$  when the layer
densities are reduced to $2\times 10^{10}cm^{-2}$. Ohmic contacts are  placed at the
ends of four arms extending   outward from the central mesa.  Using a selective
depletion scheme\cite{technique}, these  contacts can  be connected to both 2D layers
in parallel or to either layer individually.  Consequently,  both conventional
resistivity and interlayer tunneling measurements can be made on the  same sample
during a single cooldown from room temperature.  Qualitatively identical  tunneling
data were obtained from three distinct samples taken from the same MBE wafer. 

At zero magnetic field and low temperature the tunneling current-voltage
($I$-$V$)  characteristics of our samples exhibit the simple sharp resonances
characteristic of a high  degree of momentum and energy conservation upon
tunneling\cite{murphy2}.  The tunneling  conductance $dI/dV$ exhibits a sharp peak
which, for equal layer densities, is centered  at zero interlayer bias voltage $V$. 
The width of this peak ranges from 0.15meV for  $N_1=N_2$=$5.4 \times 10^{10}cm^{-2}$
to about 0.2meV at $2.1\times10^{10}cm^{-2}$  and reflects the static disorder in the
system\cite{murphy2}. The peak conductance is  typically only about $3 \times 10^{-8}
\Omega^{-1}$. The samples were designed to be weakly tunneling in order
to avoid problems arising from the small sheet  conductivities of the 2D layers which
develop at high magnetic field.

\begin{figure}
\begin{center}
\epsfxsize=3.3in
\epsfclipon
\epsffile{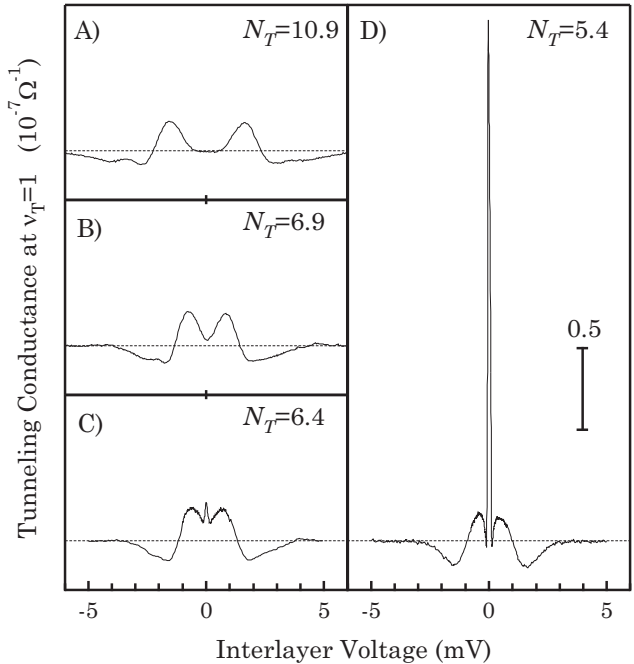}
\end{center}
\label{fig1}
\caption[figure 1]{
Tunneling conductance $dI/dV$ vs. interlayer voltage $V$ at $\nu_T=1$ and T=40mK in a
balanced double layer 2D electron system. Each trace corresponds to a different total
density $N_T$ (in units of $10^{10}cm^{-2})$,  and thus a different magnetic field. 
Trace A, at the highest  density, shows a deep suppression of the tunneling near zero
bias.  By trace D, the lowest  density of the four shown, this suppression has been
replaced by a tall peak. The vertical  scale is the same for all traces.
}
\end{figure}

Figure 1 displays the central result of this investigation.  Four low temperature 
(T=40mK) tunneling conductance spectra are shown, each at a different total density
$N_T$  in the bilayer system.   In each case the individual layer densities were
carefully matched  by adjusting (via the gates) the symmetry and voltage location of
the tunnel resonance at  zero magnetic field.  The traces were taken at different
magnetic fields but each  represents total Landau level filling factor $\nu_T$=1.  For
trace A the density is relatively  high, $N_T=10.9 \times 10^{10}cm^{-2}$, and the
well-known coulombic suppression of tunneling at the  Fermi level (i.e. at $V$=0) is
clearly evident\cite{jpe2,brown}. Trace B, taken at $N_T=6.9 \times 10^{10}cm^{-2}$, is
similar to A except that the suppression effect is weaker and the overall spread in
energy of the  tunneling is less.  This is the expected behavior since the
inter-particle Coulomb energy  falls with density.   Trace C, at $N_T=6.4 \times
10^{10}cm^{-2}$,  reveals a  qualitatively new feature: a small yet sharp peak in
$dI/dV$ at $V=0$. Finally, at the  still lower density $N_T=5.4 \times 10^{10}cm^{-2}$,
trace D shows that the peak has become  enormous and dwarfs all other features in the
tunnel spectrum.  The height of this peak  continues to grow as the density is reduced
to $3.2 \times 10^{10}cm^{-2}$ where it exceeds  even the zero magnetic field tunneling
conductance by more than a factor of 10\cite{sigmaxx}.  

It is important to note that while the two layers have equal densities at $V=0$, the
finite  interlayer capacitance disrupts this balance at non-zero $V$.  For the data in
Fig. 1, this  effect preserves $\nu_T=1/2+1/2=1$ since carriers are merely shifted from
one layer to the other.  We have found it possible to compensate for this effect, i.e.
maintain the individual layers  at fixed density, by adjusting the top and bottom gate
voltage in linear proportion to the  interlayer voltage $V$.  This compensation alters
the details of the tunneling characteristic  at large $V$, but it has a negligible
effect near $V=0$.  Since our primary focus is the resonant  peak at $V=0$, we shall
for simplicity restrict the subsequent discussion to data obtained  without
compensating for the capacitive charge transfer.

\begin{figure}
\begin{center}
\epsfxsize=3.3in
\epsfclipon
\epsffile{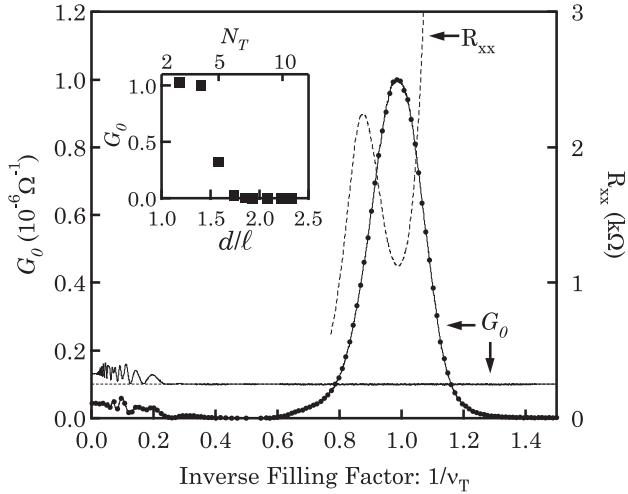}
\end{center}
\label{fig2}
\caption[figure 2]{
Zero bias tunneling conductance $G_0$ vs. inverse filling factor at high and low
density. Light solid trace: $N_T=10.9\times 10^{10}cm^{-2}$; data displaced vertically for
clarity. Above about  $\nu_T^{-1}=0.3$ the conductance is nearly zero. Dotted trace:
$N_T=4.2\times 10^{10}cm^{-2}$.  A huge  enhancement of the tunneling is observed near
$\nu_T=1$. Dashed curve: Longitudinal resistance  $R_{xx}$ for $N_T=4.2\times
10^{10}cm^{-2}$ showing QHE minimum at $\nu_T=1$.  Inset: Tunneling  conductance at
$\nu_T=1$ vs. $d/\ell$.  All data taken at T=40mK.
}
\end{figure}

Figure 2 shows the magnetic field dependence of the zero bias (i.e. $V=0$) tunneling 
conductance $G_0$, at T=40mK, for $N_T=10.9 \times 10^{10}$ and $4.2 \times
10^{10}cm^{-2}$. In order to directly  compare the two curves, the data is plotted against
the inverse total filling factor  $\nu_T^{-1}=eB/hN_T$, instead of magnetic field.  As
expected, both curves exhibit quantum  oscillations of the tunneling conductance at small
magnetic field.  These oscillations are  less pronounced in the low density data owing to
the reduced electron mobility. At high  magnetic field, in the vicinity of $\nu_T$=1, the
two data sets differ qualitatively.   The high  density data show that for magnetic fields
above about $\nu_T^{-1}$$\approx$$0.3$, the tunneling  conductance is near zero.  This
again is the coulombic suppression characteristic of  tunneling between two weakly coupled
2D electron systems at high magnetic field.   By  contrast, the low density data show an
enormous enhancement of the tunneling around  $\nu_T=1$.  The enhancement appears
strongest at, or very near, $\nu_T=1$, but it is  clearly a very substantial effect over a
wide range of filling factors.  

The temperature dependence of the zero bias tunneling conductance is displayed in Fig. 
3.  Two sets of data are shown, one for $N_T=10.9 \times 10^{10}cm^{-2}$ and one for
$N_T= 4.2 \times 10^{10}cm^{-2}$. There is again a qualitative difference  between the
tunneling at high and low density.  At high density the conductance falls  with
decreasing temperature.  As reported previously, this dependence is consistent  with
simple thermal activation\cite{jpe2}.  The low density data behave in the  opposite
fashion, rising as the temperature falls.  The rise becomes fairly  steep around T=0.4K
and then levels off below about 40mK.

Magneto-transport measurements on similar double quantum well samples have  established
the approximate location of the phase boundary between the incompressible  $\nu_T=1$
quantized Hall phase and the compressible non-quantized Hall phase at large layer 
separation\cite{murphy}.  In the limit of weak tunneling this boundary was found to be 
near $d/\ell\approx2$. In this ratio $d$ is the center-to-center distance between the 
quantum wells and $\ell=(\hbar/eB)^{1/2}$ is   the magnetic length. In its as-grown
state, i.e. when $N_T=10.9 \times 10^{10}cm^{-2}$,  the present sample has $d/\ell=2.4$
at $\nu_T=1$.  Reducing the density via gating to  $N_T= 4.2 \times 10^{10}cm^{-2}$,
gives $d/\ell=1.45$.  Control over the layer densities  thus allows us to span the
expected phase boundary using a single sample.  The inset  to Fig. 2 shows the zero
bias tunneling conductance $G_0$ at $\nu_T=1$, at T=40mK,  versus $d/\ell$.  There is a
sharp transition near $d/\ell\approx1.8$ separating  two very different tunneling
regimes.  For $d/\ell>1.8$ the zero bias conductance  is suppressed and the tunneling
spectra are qualitatively the same as seen in  samples having negligible interlayer
correlations.  On the other hand, as $d/\ell$ falls below this critical value a
resonant enhancement of the tunneling appears at zero bias.  The magnitude of this
peak grows continuously as $d/\ell$ falls\cite{sigmaxx}.

\begin{figure}
\begin{center}
\epsfxsize=3.3in
\epsfclipon
\epsffile{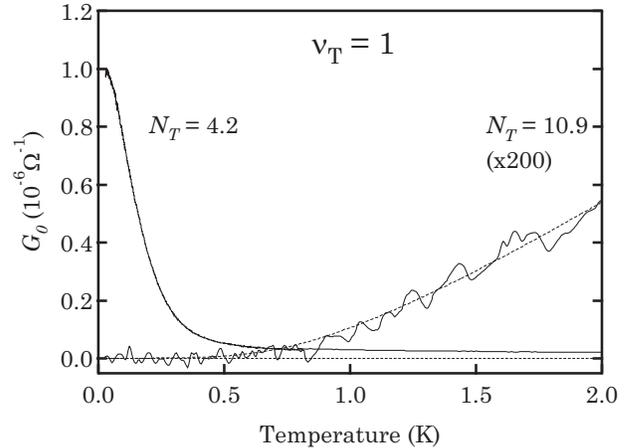}
\end{center}
\label{fig3}
\caption[figure 3]{
Temperature dependence of the zero bias tunneling conductance at $\nu_T=1$ at high and
low densities. Note that the high density data has been magnified by a factor of 200.
}
\end{figure}

The rough agreement between the critical $d/\ell$ value found in transport 
experiments\cite{murphy} and that reported here in tunneling studies suggests  that they
reflect the {\it same} phase transition. To investigate this, resistivity  measurements
were performed on the sample.   As  anticipated, a quantized Hall effect does develop at
low density.  The dotted trace in Fig.  2 shows the observed minimum in $\rho_{xx}$ at
$N_T=4.2\times10^{10}cm^{-2}$ and T=40mK.   The relative weakness of this many-body
integer QHE may simply be due to the low mobility  of the sample
($\sim2\times10^{5}cm^2/Vs$) at this reduced density.  A second possibility is more
interesting: It is well known that for some interaction-driven quantized Hall states 
Arrhenius behavior of the resistivity $\rho_{xx}$ is not observed until the temperature
is  reduced well below the measured activation energy.  For example, in conventional
single layer  2D systems at $\nu=1$, spin flip activation energies of order 40K are
common, but can only  be observed at temperatures below a few Kelvin\cite{schmeller}. 
This effect originates in  the low energy of long wavelength spin waves in the system. 
The activation energy measured  in transport reflects the large exchange contribution to
the energy required to create a  well-separated quasiparticle-quasihole pair.  By
contrast, at long wavelengths the collective  mode energy is the much smaller bare
Zeeman energy.  Thermal population of large numbers  of these modes is apparently
effective in suppressing the onset of a thermally  activated resistivity.  The same
phenomenon is observed in the $\nu_T=1$ QHE in double layer  systems\cite{murphy,lay}. 
Now, however, the collective modes are the predicted pseudospin  waves.  At $q=0$ these
modes are gapless, but only in the non-physical limit of zero tunneling.  In real 
samples a gap at $q=0$ is expected, the size of which is determined by both the single 
particle tunneling energy $\Delta_{SAS}$ and an interlayer capacitive charging 
energy\cite{wen,yang}.  Using our estimate\cite{dsas} of $\Delta_{SAS}=90 \mu K$  and
published\cite{yang} estimates of the charging energy, the long wavelength pseudospin 
wave energy is only about 70mK.  It is therefore not so surprising that the $\rho_{xx}$ 
minimum at $\nu_T=1$ is weak at 40mK and absent above about 200mK. We note in passing
that  the estimated $\Delta_{SAS}$ in our sample is only $1.3\times10^{-6}$ of the
typical Coulomb  energy $e^2/\epsilon\ell$ at $N_T=4.2\times10^{10}cm^{-2}$.  This is by
far the smallest tunneling  strength of any sample reported exhibiting the $\nu_T=1$
bilayer QHE state. 

A qualitative explanation for the tunneling enhancement reported here can be
constructed from  known aspects of double layer systems at $\nu_T=1$.  At high
densities, when the layers are  weakly coupled, the zero bias tunneling is heavily
suppressed.  The energetic penalty associated with tunneling in this case arises
because an electron attempting to tunnel is totally "unaware" of the strong
correlations present within the layer it is about to enter.  In the $\nu_T=1$  bilayer
QHE state just the opposite is true; the very essence of the state lies in the strong 
interlayer correlation it contains.  Indeed, the so-called $\Psi_{111}$
wavefunction\cite{book,chak} believed to represent this  QHE state may be viewed as a
collection of correlated interlayer excitons.  An electron in  one layer is always
opposite a hole in the other layer.  It is not implausible that this  interlayer
correlation enhances tunneling.  This simplistic view, while appealing,  does not
address the sharply resonant nature of the enhancement.  The experimental observation 
of a narrow peak in the tunneling conductance at zero interlayer voltage suggests that 
there is a collective mode near zero energy which can transfer charge between the two
layers.   The predicted pseudospin Goldstone mode does exactly this.  This mode
involves oscillations  of the pseudospin magnetization both in the $xy$-plane and in
the $z$-direction of pseudospin  space. This latter component resembles an antisymmetric 
interlayer plasma oscillation. In the absence of tunneling no charge is transferred between 
the layers and the mode energy vanishes in the long wavelength, $q=0$ limit.  In a real 
sample however, the finite tunneling amplitude leads to an energy gap at $q=0$ and, most 
importantly, allows the collective mode to transfer charge. The estimated gap 
($\sim 70$mK=6$\mu$eV) is much smaller than the observed width of the tunnel resonance 
(typically 150$\mu$eV at low temperature) so it is not surprising that a single peak in 
dI/dV appears at V=0.  From this point of view, tunneling appears to provide direct 
spectroscopic evidence for the Goldstone mode of the broken symmetry QHE state at $\nu_T=1$.

In conclusion, we have examined tunneling in double layer 2D electron systems in which 
interlayer correlations are very strong.  A dramatic resonant enhancement of the zero
bias  tunneling conductance is observed when the system crosses the phase boundary from
a  compressible fluid into the $\nu_T=1$ ferromagnetic quantized Hall state.  Although
detailed  understanding of the tunnel spectra is lacking, it seems likely that the
observed resonance  is intimately connected with the Goldstone mode of the broken
symmetry state.

We are indebted to A.H. MacDonald, S.M. Girvin, and A.C. Gossard for
numerous useful discussions and to the National Science Foundation and
the Department of Energy for financial support. One of us (I.B.S.)
acknowledges the Department of Defense for a National Defense Science
and Engineering Graduate Fellowship.


\begin{references}
\bibitem{sigmaxx} At the lowest densities the tunneling conductance at $\nu_T=1$ 
becomes so large that it may be slightly affected by the finite sheet conductivity 
of the 2D system.
\bibitem{dsas}This is a rough estimate based on a self-consistent calculation of 
the energy levels in the DQW.  Simple $\Gamma$-point tunneling and a uniform
effective mass of $0.067m_0$ were assumed.
\end{references}
\end{document}